# Spin-pumping-induced inverse spin-Hall effect in Nb/Ni$_{80}$Fe$_{20}$ bilayers and its strong decay across the superconducting transition temperature

Kun-Rok Jeon,[1,2] Chiara Ciccarelli,[2] Hidekazu Kurebayashi,[3] Jöerg Wunderlich,[4,5] Lesley F. Cohen,[6] Sachio Komori,[1] Jason W. A. Robinson,[1] and Mark G. Blamire[1]

[1]*Department of Materials Science and Metallurgy, University of Cambridge, 27 Charles Babbage Road, Cambridge CB3 0FS, UK*

[2]*Cavendish Laboratory, University of Cambridge, Cambridge CB3 0HE, UK*

[3]*London Centre for Nanotechnology and Department of Electronic and Electrical Engineering at University of College London, London WC1H 01H, UK*

[4]*Hitachi Cambridge Laboratory, J. J. Thomson Avenue, Cambridge CB3 0HE, UK*

[5]*Institute of Physics, ASCR, Cukrovarnicka 10, 162 00 Praha 6, Czech Republic*

[6]*The Blackett Laboratory, Imperial College London, SW7 2AZ, UK*

**We quantify the spin Hall angle $\theta_{SH}$ and spin diffusion length $l_{sd}$ of Nb from inverse spin-Hall effect (iSHE) measurements in Nb/Ni$_{80}$Fe$_{20}$ bilayers under ferromagnetic resonance. By varying the Nb thickness $t_{Nb}$ and comparing to a Ni$_{80}$Fe$_{20}$/Pt reference sample, room temperature values of $\theta_{SH}$ and $l_{sd}$ for Nb are estimated to be approximately $-0.001$ and 30 nm, respectively. We also investigate the iSHE as a function of temperature $T$ for different $t_{Nb}$. Above the superconducting transition temperature $T_c$ of Nb, a clear $t_{Nb}$-dependent $T$ evolution of the iSHE is observed whereas below $T_c$, the iSHE voltage drops rapidly and is below the sensitivity of our measurement setup at a lower $T$. This suggests the strong decay of the quasiparticle (QP) charge-imbalance relaxation length across $T_c$, as supported by an additional investigation of the iSHE in a different sample geometry along with model calculation. Our finding suggests careful consideration should be made when developing superconductor spin-Hall devices that intend to utilize QP-mediated spin-to-charge interconversion.**

# I. INTRODUCTION

The flow of spin angular momentum without an accompanying net charge current, so-called pure spin current, is a key ingredient of spintronic devices mostly consisting of ferromagnet (FM)/non-magnet (NM) heterostructures. This pure spin current enables us to transmit spin information through the NM with low energy dissipation and to control the magnetization $M$ of the FM via spin transfer torque [1-5]. It has been well-established that ferromagnetic resonance (FMR) spin pumping [6,7], the dynamic transfer of spin angular momentum from a precessing FM into an adjacent NM, can provide an attractive and powerful method for generating the pure spin current.

The combination of FMR spin pumping with inverse spin Hall effect (iSHE) [8-10], spin-to-charge conversion, allows for the electrical detection of the generated spin currents in a FM/NM bilayer. A dynamically injected spin current $J_s$ in the NM layer is converted into a transverse charge current $J_c$ via the iSHE, producing a measurable electromotive force [Fig. 1(a)]. This approach has been widely employed to investigate the spin-orbit coupling and spin transport parameters, such as spin Hall angle $\theta_{SH}$ and spin diffusion length $l_{sd}$, in a variety of NM materials, including metals [9], semiconductors [11,12], oxide interfaces [13,14], and topological insulators [15,16].

Recent progress in superconducting spintronics [17,18] has highlighted the potential of superconductors (SCs) towards future low-energy computing technologies. Several studies exploring the quasiparticle (QP) spin transport in SCs have been achieved using DC (non-)local transport measurements [18-25]. Interestingly, it has been shown that in all metallic non-local spin-Hall devices with transparent contacts [25], the QP-mediated iSHE in the superconducting state of NbN increases significantly by about 3 orders of magnitude compared to that in the normal state. Another recent experiment has reported that for a ferrimagnetic insulator YIG/NbN junction with ohmic contacts [26], the iSHE voltage induced by the spin Seebeck effect is enhanced by a factor of ~2.5 in the vicinity of the superconducting transition. Although more work is certainly needed, these experiments seem to suggest the existence of emergent phenomena arising through QP spin-orbit coupling. This motivates us to investigate the QP-mediated iSHE in Nb, the standard material for superconducting electronics and spintronics.

Here, we experimentally quantify the $\theta_{SH}$ and $l_{sd}$ values of Nb films from spin-pumping-induced iSHE measurements in Nb/Ni$_{80}$Fe$_{20}$ bilayers by varying the Nb thickness $t_{Nb}$ and comparison with a Ni$_{80}$Fe$_{20}$/Pt reference sample. Spin precession effect under an oblique magnetic field also enables a first-order estimate of the spin lifetime in the Nb. Furthermore, we study the iSHE as a function of temperature $T$ for different $t_{Nb}$. Above the superconducting transition temperature $T_c$ of Nb, a clear $t_{Nb}$-dependent $T$ evolution of the iSHE is observed. Yet below $T_c$, the iSHE voltage drops rapidly and becomes unmeasurable at a lower $T$, which can be explained by the short QP charge-imbalance relaxation length in the superconducting Nb. Our experiments along with model calculation suggest the necessity of a careful design of the sample/device geometry in spin-pumping-induced iSHE measurements with SCs below $T_c$.

## II. EXPERIMENTAL DETAILS

We prepared Nb/Ni$_{80}$Fe$_{20}$ structures, Ni$_{80}$Fe$_{20}$/Nb inverted structures, and Pt/Ni$_{80}$Fe$_{20}$ reference samples on either thermally oxidized Si or quartz substrates with lateral dimension of 3−5 mm × 5 mm by dc magnetron sputtering in an ultra-high vacuum chamber. Note that the Ni$_{80}$Fe$_{20}$/Nb inverted structures were used for the study of the sample geometry dependence by simplifying the patterning process. While $t_{Nb}$ ranges from 7.5 to 60 nm, the Ni$_{80}$Fe$_{20}$ (Pt) thickness is fixed at 6 nm (5 nm). Details of the sample preparation can be found elsewhere [27]. The $T_c$ of the Nb layers was determined by DC electrical transport measurements (see Ref. [28]). Hereafter, $T_c$ denotes the value determined under microwave excitation unless otherwise specified. Single-stripe-patterned samples were prepared by conventional microfabrication techniques (e.g. photo-lithography, Ar-ion beam etching).

The measurement setup used for this study [Fig. 1(a)] is based on broad-band FMR techniques [27]. The sample was attached face-down on the coplanar waveguide (CPW) by using an electrically insulating high-vacuum grease. A MW signal was passed through the CPW and excited FMR of the Ni$_{80}$Fe$_{20}$ layer; a transverse DC voltage as a function of external static magnetic field was measured between two Ag-paste contacts at opposite ends of the sample. *Simultaneously*, we measured the absorbed MW power where the FMR was excited. We employed a vector field cryostat from *Cryogenic Ltd* that allows for a 1.2 T magnetic field in any direction over a wide $T$ range of 2−300 K.

## III. RESULTS AND DISCUSSION

### A. Nb thickness dependence of inverse spin-Hall effect in Nb/Ni$_{80}$Fe$_{20}$ bilayers

We start by describing the spin-pumping-induced iSHE in Nb/Ni$_{80}$Fe$_{20}$ samples at 300 K. Figure 2 shows the FMR absorption (top panel) and transverse DC voltage measurements (bottom panel) vs. external magnetic field $\mu_0 H$ along the $x$-axis for three different $t_{Nb}$ (7.5, 30, and 60 nm). In these measurements, the MW frequency was fixed at 5 GHz and the MW power at the CPW at ~100 mW. In all the samples, the FMR of the Ni$_{80}$Fe$_{20}$ is excited around the resonance magnetic field $\mu_0 H_{res}$ and a clear Lorentzian peak emerges in the DC voltage. Importantly, the polarity of the Lorentzian peak is inverted by reversing the magnetic field, which is consistent with the symmetry of iSHE [8-10].

The measured (DC) voltage can be decomposed into symmetric and anti-symmetric Lorentzian functions with respect to $\mu_0 H_{res}$, with weights of $V_{sym}$ and $V_{asy}$ respectively:

$$V(H) = V_{sym}(H) + V_{asy}(H) + V_0,$$

$$V_{sym}(H) = V_{sym} \cdot \left[ \frac{(\Delta H)^2}{(\Delta H)^2 + (H - H_{res})^2} \right], \quad V_{asy}(H) = V_{asy} \cdot \left[ \frac{(\Delta H) \cdot (H - H_{res})}{(\Delta H)^2 + (H - H_{res})^2} \right], \quad (1)$$

where $V_0$ is a background voltage. All the data are well fitted by Eq. (1). We note that in principle, $V_{sym}$ is attributed not only to the iSHE but also to the spin rectification effect (SRE) [29-31]. However, in our setup the iSHE contribution turns out to be predominant, as discussed in more detail below.

A typical MW power ($P_{MW}$) dependence of $V_{sym}$, extracted from the data $t_{Nb} = 7.5$ nm [Fig. 2(d)], is shown in Fig. 2(e). The extracted $V_{sym}$ scales almost linearly with $P_{MW}$, as expected for the FMR spin pumping in linear response regime ($J_s \propto P_{MW}$) [8-10]. To check the sign of $\theta_{SH}$ in Nb, we repeated the same measurement on a Pt/Ni$_{80}$Fe$_{20}$ reference sample [Fig. 2(f)], where the Pt is well known to have a positive $\theta_{SH}$ [8,9,31]. Opposite signs of $V_{sym}$ are observed in the Nb and Pt spin sink samples [Figs. 2(a) and 2(f)], confirming the negative $\theta_{SH}$ of Nb [24,33]. Moreover the sign change in $V_{sym}$ indicates that the iSHE, rather than the SRE [8-10], gives a dominant contribution to $V_{sym}$.

To quantify the spin Hall angle $\theta_{SH}$ and the spin diffusion length $l_{sd}$ in the Nb films, we plotted the effective Gilbert damping $\alpha$ [Fig. 3(a)] and $V_{sym}$ [Fig. 3(b)] as a function of $t_{Nb}$. The values of $\alpha$ and the effective saturation magnetization $\mu_0 M_{eff}$ [inset of Fig. 3(a)] were deduced from the MW frequency $f$ dependence of FMR spectra (e.g. the FMR linewidth $\mu_0\Delta H$ and the resonance field $\mu_0 H_{res}$, see Ref. [28]). The $t_{Nb}$-dependent $\alpha$ enhancement, resulting from FMR spin pumping into the Nb layer [6,7], can be expressed by

$$\alpha(t_{SC}) = \alpha_0 + \alpha_{sp}(t_{SC}),$$

$$\alpha_{sp}(t_{SC}) = \left(\frac{g_L\mu_B g_r^{\uparrow\downarrow}}{4\pi M_s t_{FM}}\right) \cdot \left[1 + \frac{g_r^{\uparrow\downarrow}\mathcal{R}_{SC}}{\tanh\left(\frac{t_{SC}}{l_{sd}^{SC}}\right)}\right]^{-1}, \quad (2)$$

where $\alpha_0$ and $\alpha_{sp}$ are, respectively, the FMR damping irrelevant and relevant to the spin pumping, $g_L$ is the Landé g-factor taken to be 2.1 [35], and $\mu_B$ is the Bohr magneton. $g_r^{\uparrow\downarrow}$ is the effective real-part spin-mixing conductance across a Nb/Ni$_{80}$Fe$_{20}$ interface. $\mathcal{R}_{SC} \equiv \rho_{SC} l_{sd}^{SC} e^2/2\pi\hbar$ is the spin resistance, $\rho_{SC}$ is the resistivity of the Nb [inset of Fig. 3(b)], and $e$ is the electron charge. $t_{FM}$ and $t_{SC}$ are the Ni$_{80}$Fe$_{20}$ thickness (6 nm) and the Nb thickness (7.5 − 60 nm), respectively. Fitting Eq. (2) to $\alpha(t_{Nb})$ [blue line in Fig 3(a)] yields $g_r^{\uparrow\downarrow} = 16 \pm 3$ nm$^{-2}$ and $l_{sd}^{SC} = 35 \pm 2$ nm at 300 K. The estimated $l_{sd}^{SC}$ is in the same range of reported previously for Ni$_{80}$Fe$_{20}$/Nb/Ni$_{80}$Fe$_{20}$ spin valves [34].

By combining the calculated spin current density $j_s$ at the Nb/Ni$_{80}$Fe$_{20}$ interface with the measured $V_{sym}$ (or charge current $I_c$) [Fig. 3(b)], one can estimate the spin-to-charge conversion efficiency parameterized by $\theta_{SH}$:

$$j_s \approx \left(\frac{G_r^{\uparrow\downarrow}\hbar}{8\pi}\right) \cdot \left(\frac{\mu_0 h_{rf}\gamma}{\alpha}\right)^2 \cdot \left[\frac{\mu_0 M_{eff}\gamma + \sqrt{(\mu_0 M_{eff}\gamma)^2 + 16(\pi f)^2}}{(\mu_0 M_{eff}\gamma)^2 + 16(\pi f)^2}\right] \cdot \left(\frac{2e}{\hbar}\right), \quad (3)$$

$$V_{iSHE} = \left(\frac{R_{FM}R_{SC}}{R_{FM} + R_{SC}}\right) \cdot I_c = \left(\frac{w_y}{\sigma_{FM}t_{FM} + \sigma_{SC}t_{SC}}\right) \cdot \theta_{SH} l_{sd}^{SC} \cdot \tanh\left(\frac{t_{SC}}{2l_{sd}^{SC}}\right) \cdot j_s, \quad (4)$$

where $G_r^{\uparrow\downarrow} \equiv g_r^{\uparrow\downarrow} \cdot \left[1 + g_r^{\uparrow\downarrow}\mathcal{R}_{SC}/\tanh\left(\frac{t_{SC}}{l_{sd}^{SC}}\right)\right]^{-1}$. $\gamma = g_L\mu_B/\hbar$ is the gyromagnetic ratio of $1.84 \times 10^{11}$ T$^{-1}$ s$^{-1}$ and $\hbar$ is Plank's constant divided by $2\pi$. $\mu_0 h_{rf}$ is the amplitude of MW magnetic field (0.2 mT for 100 mW) [36]. $R_{FM}(R_{SC})$ and $\sigma_{FM}(\sigma_{SC})$ are the square resistance and the conductivity of the Ni$_{80}$Fe$_{20}$ (Nb) layer [inset of Fig. 3(b)], respectively. $w_y$ is the width of MW transmission line (1 mm, see Fig. 1) for the un-patterned samples. From the data in Fig. 3(b) using $g_r^{\uparrow\downarrow} = 16 \pm 3$ nm$^{-2}$ and Eq. (4), we obtain the room temperature (RT) values of $\theta_{SH} \approx -0.001$ and $l_{sd}^{SC} \approx 30$ nm for the Nb film. This $\alpha_{SH}$ value, corresponding to the spin Hall conductivity $\sigma_{SHE} \approx -0.06 \times 10^3$ $\Omega^{-1}$-cm$^{-1}$, is in good agreement with that expected from theoretical calculations [37]. We also note that in a previous experiment of the non-local spin valve with a rather resistive Nb ($\rho_{Nb} = 90$ μΩ-cm at 10 K), a larger $\theta_{SH}$ of $-0.009$ and a smaller $l_{sd}^{SC}$ of 6 nm were obtained [33], giving $\sigma_{SHE} = -0.10 \times 10^3$ $\Omega^{-1}$-cm$^{-1}$. This value is similar to what we obtained.

## B. Out-of-plane angular dependence and oblique Hanle spin precession

We measure the out-of-plane angular dependence of DC voltages [Fig. 4(a)] to extrapolate the spin lifetime $\tau_{sf}$ in Nb. The results discussed here corroborate that the observed $V_{sym}$ signals are ascribed to the spin-pumping-induced iSHE in the Nb layer. When $\mu_0 H$ is applied at an angle $\theta_H$ to the $x$-axis [inset of Fig. 4(a)], the angle $\phi_M$ of $M$ precession axis does not necessarily coincide with $\theta_H$ because of the demagnetization energy (or shape anisotropy energy). The corresponding misalignment angle $(\theta_H - \phi_M)$ on FMR is given by [38]

$$(\theta_H - \phi_M) \approx \arctan\left[\text{sgn}(\theta_H) \cdot \sqrt{\left(\frac{\cos(2\theta_H) + (\mu_0 H_{res}/\mu_0 M_{eff})}{\sin(2\theta_H)}\right)^2 + 1} - \frac{\cos(2\theta_H) + (\mu_0 H_{res}/\mu_0 M_{eff})}{\sin(2\theta_H)}\right]. \quad (5)$$

The $\theta_H$ dependence of $\phi_M$, calculated from Eq. (5) with the measured value of $\mu_0 H_{res}$ [Fig. 4(b), top panel], is shown in the inset of Fig. 4(b). This misalignment $(\theta_H - \phi_M)$ can give rise to the Hanle effect [39], in which the static $\mu_0 H$ transverse to the pumped spins $S(t)$ suppresses the spin accumulation in the spin sink via spin precession and dephasing [inset of Fig. 4(a)], if $\tau_{sf}$ is comparable to or longer than the Larmor precession time $1/\omega_L$. This results in the characteristic angular dependence of the voltage signal [40,41]:

$$V_{iSHE}(\theta_H) \propto \left\{\cos(\theta_H) \cdot \cos(\theta_H - \phi_M) + \sin(\theta_H) \cdot \sin(\theta_H - \phi_M) \cdot \left[\frac{1}{1 + (\omega_L \cdot \tau_{sf})^2}\right]\right\} \quad (6)$$

with $\omega_L = g_L \mu_B \cdot (\mu_0 H)/\hbar$ is the Larmor frequency. It is worth noting that in the case of a short $\tau_{sf}$ [red symbol in Fig. 4(b)], $V_{iSHE}(\theta_H)$ is simply proportional to $\cos(\phi_M)$. On the other hand, if $\tau_{sf}$ increases [$\geq 1/\omega_L$, black and blue symbols in Fig. 4(b)], the Hanle spin precession effectively reduces $V_{iSHE}(\theta_H)$ in particular around $\theta_H = 80°$, where the absolute of $(\theta_H - \phi_M)$ is maximun [upper inset of Fig. 4(b)]. The measured $V_{sym}(\theta_H)$ in the Nb/Ni$_{80}$Fe$_{20}$ bilayer is fairly reproduced by Eq. (6) with $\tau_{sf}$ of the order of a few ps [lower inset of Fig. 4(b)]. This is also consistent with the estimated value of 2−3 ps using $\tau_{sf}^{SC} = (l_{sd}^{SC})^2/D_{SC}$ with $D_{SC}$ is the diffusion coefficient of Nb (10−15 cm$^2$/s at RT) and $l_{sd}^{SC} \approx 30$ nm obtained from $V_{sym}(t_{Nb})$ [Fig. 3(b)]. The iSHE in a Ni$_{80}$Fe$_{20}$ layer could, in principle, contribute to $V_{iSHE}(\theta_H)$ [42]. However, $\tau_{sf} = 0.025$ ps in the Ni$_{80}$Fe$_{20}$ calculated using $D_{FM} = 10$ cm$^2$/s and $l_{sd}^{FM} = 5$ nm [43] is too short ($<< 1/\omega_L \approx 8$ ps for $\mu_0 H_{res} = 0.7-0.8$ T around $\theta_H = 80^o$) to cause the noticeable suppression of $V_{iSHE}$. This result further confirms that the measured $V_{sym}$ signals in our system originate from the spin-pumping-induced iSHE in the Nb layer.

### C. Temperature evolution of spin-pumping-induced inverse spin-Hall effect

Next, we investigate the $T$ dependence of $V_{sym}$ for the Nb/Ni$_{80}$Fe$_{20}$ samples with three different $t_{Nb}$ of 7.5, 30, and 60 nm [Fig 5(a)]. As summarized in Fig. 5(b), for $t_{Nb} = $ 7.5 nm (non-superconducting down to 2 K), $V_{sym}$ is visible in the entire $T$ range, varying slightly as $T$ decreases. In contrast, for the thicker superconducting samples ($t_{Nb} = 30$ nm and 60 nm), $V_{sym}$ is reduced gradually with decreasing $T$ from 300 to 10 K. When $T < 8$ K (entering the superconducting state), the voltage signal drops abruptly and becomes below the sensitivity of our measurement setup at a lower $T$. The $t_{Nb}$-dependent $T$ evolution of $V_{sym}$ in the normal state is qualitatively understood in terms of the $t_{Nb}$-dependent $T$ evolution of $\rho_{Nb}$ [inset of Fig. 4(d)] and $G_r^{\uparrow\downarrow}$ [see Eqs. (3) and (4)]. Note that the trade-off of the $\rho_{Nb}$ reduction and the $G_r^{\uparrow\downarrow}$ enhancement with decreasing $T$ determines the overall $T$ dependence of $V_{iSHE}$. In our system, we observed no clear signature of the coherence effect of superconductivity (see Ref. [28] for detailed data), namely, anomalous enhancement of spin current flow near $T_c$ that results from the well-developed coherence peaks of the SC density of states being accessible to the spin-transporting QPs [26,44,45]. This supports the previous studies [44-46] that for a *metallic/conducting* FM in direct contact with SC, $\Delta$ is significantly suppressed at the FM/SC interface due to the (inverse) proximity effect of the FM, leading to the vanishing of the superconducting coherence peak effect [44-46]. How local $T$ increase due to MW power absorption influences the voltage signal immediately below $T_c$ is also discussed in Ref. [28].

### D. Model calculation of quasiparticle-mediated spin-Hall voltages in Nb films

To understand why the iSHE voltages (in our setup) have vanished deep into the superconducting state, we consider the decay of the charge imbalance effect caused by non-equilibrium electron-like or hole-like QP states [23,25,47,48], namely, the charge-imbalance relaxation length $\lambda_Q$. In the diffusive case, $l_{sd}$ is longer than the mean free path $l_{mfp}$ [47-49],

$$\lambda_Q = \sqrt{D_Q \tau_Q}, \quad \tau_Q \approx \frac{4 k_B T}{\pi \Delta(T)} \cdot \tau_\varepsilon, \quad (7)$$

where $D_Q = \left[2 f_0(\Delta) / \chi_Q^0(T)\right] \cdot D$ is the charge diffusion coefficient of the QPs [50,51], $f_0(\Delta) = \left[\exp(\Delta / k_B T) + 1\right]^{-1}$ is the Fermi-Dirac (FD) distribution function at $\Delta$, and $\chi_Q^0(T) = 2\int_\Delta^\infty \left(\sqrt{E^2 - \Delta^2} / E\right) \cdot \left[-\partial f_0(E) / \partial E\right] dE$ is the normalized charge susceptibility of QP [50,51]. $\tau_{qp}$ is the charge-imbalance relaxation time, $\tau_\varepsilon$ is the energy relaxation time, and $\Delta(T) \approx 1.76 k_B T_c \cdot \tanh\left[1.74 \sqrt{T / T_c - 1}\right]$ is the superconducting energy gap. Note that $k_B T / \Delta$ represents an approximate estimate for the fraction of QPs participating in the charge imbalance [47-49]. Around $T_c$ because $\tau_\varepsilon$ does not change significantly, $\lambda_Q(T) \propto \left[\Delta(T)\right]^{-1/2} \propto (1 - T / T_c)^{-1/4}$. By contrast below $T_c$, $k_B T / \Delta(T)$ is of the order of unity and this means that $\lambda_Q(T)$ is determined by $\tau_\varepsilon(T)$. If the QP charge relaxation is dominated by the inelastic electron-phonon scattering, $\tau_{in} \propto T^{-3}$ for low energy QPs [$k_B T \ll \Delta(T)$] and thus $\lambda_Q(T) \propto T^{-3/2}$ [47-49]. Considering all this, the overall $T$ dependence can be approximated by $\lambda_Q(T) \approx \lambda_Q(0) \cdot \left[T^{-3/2} + (1 - T / T_c)^{-1/4}\right]$. It was previously shown from current-voltage characteristics of Nb nanobridges [52] and spin resistance measurements in Ni$_{80}$Fe$_{20}$/Al$_2$O$_3$/Nb/Al$_2$O$_3$/Ni$_{80}$Fe$_{20}$ structures [53] that $\lambda_Q \approx$ 90-150 nm and $\tau_Q \approx$ 13-26 ps for Nb films immediately below $T_c$.

To gain further insight into the role of the factor $\lambda_Q(T)$, we calculated the transverse DC voltage $V_{iSHE}^Q$ expected from QP-mediated iSHE in the superconducting Nb layer (Fig. 6) according to the previous theoretical work [50,51], where the QP spin-Hall angle is assumed to be given by two extrinsic components of the side jump [54] and the skew scattering [55] (see Ref. [28] for details). The spin-to-charge conversion in SCs is rather complicated in that the coupling between different non-equilibrium modes (spin, charge, and energy) with Zeeman splitting [56-59] and the non-linear kinetic equations in the superconducting states [60-62], which have not been applied yet in non-equilibrium situations, should be taken into account properly. In the calculation, we mainly considered the change of the QP charge imbalance [23,25,47,48] because of the complexity.

The most important aspect of the calculations [Figs. 6(b) and 6(c)] is that the maximum $V_{iSHE}^Q$ at $d_y = 0$ depends insensitively on the active width of precessing FM, $w_y$ [see Fig. 1(c)], when $\lambda_Q$ becomes comparable to or shorter than $w_y$. Two $T$ regimes can be identified. For $T > T_c$, $V_{iSHE}$ scales linearly with $w_y$, as expected for the electromotive force in the normal state [8-10]; for $T < T_c$, $V_{iSHE}^Q$ is almost independent of $w_y$. We note that in addition to the rapid decay of $\lambda_Q(T)$ across $T_c$, the effective spin transport length $l_Q^*(T)$ [Fig. 6(a), middle panel] and the the QP current density $j_s^Q(T)$ [Fig. 6(a), bottom panel] are both progressively reduced as $T$ decreases due to the development of the (singlet) superconducting gap and the freeze-out of the QP population [20,25]. Thus a vanishingly small amplitude of $V_{iSHE}^Q$ [<< 1 nV, Fig. 6(b)] is expected below $T_c$ although there exists the clear rise in $V_{iSHE}^Q$ at a lower $T$, caused by the increased Nb/Ni$_{80}$Fe$_{20}$ bilayer resistance due to the exponential $T$ dependence of QP resistivity [20,25].

Notwithstanding, the calculation suggests a device geometry more suited to electrical detection of the iSHE in *both* the normal and deep into the superconducting states, namely, 1) by utilizing an array of densely-packed FM stripes with a periodicity that is comparable to the QP charge relaxation length of the SC and 2) by reducing the separation distance between the nearest FM stripes as much as possible. In such a proposed device, one can greatly amplify the total magnitude of spin-Hall voltage by increasing the active volume of QP charge imbalance for a given reasonable $P_{MW}$. Importantly, from the measured value of $V_{sym}$ = 50−150 nV (see Fig. 6), we get $V_{iSHE}^Q$ of the order of 10−100 nV, which can be *measurable* well below $T_c$. Detailed calculations are presented in Ref. [28]

### E. Sample geometry dependence of inverse spin-Hall voltages

Finally, we investigate the sample geometry dependence of iSHE voltages by using single-stripe-patterned samples to check validity of the model calculation. These samples consist of an un-etched Ni$_{80}$Fe$_{20}$/superconducting Nb bilayer at the middle and etched non-superconducting Nb leads (< 7.5 nm) on the lateral sides of the bilayer [Figs. 7(a) and 7(b)]. We note that in such patterned samples, $d_y$ can effectively be reduced to a few tens of nm, as probed by scanning electron microscope [Fig. 7(c)]. Figures 7(d)-7(g) exhibit the representative data of FMR absorption (top panel) and DC voltage measurements (bottom panel) vs. $\mu_0 H$ along the $x$-axis for two different $w_y$ of 150 and 500 μm, taken above and well below $T_c$. In the normal state ($T > T_c$), $V_{sym}$ of $w_y= 500$ μm is approximately 3 times greater than of $w_y= 150$ μm, as in accordance with the model calculation, whereas in the superconducting state ($T < T_c$), no voltage signal is observed for both cases. It is notable that the sign of $V_{sym}$ above $T_c$ is reversed from the preceding experiment with Nb/Ni$_{80}$Fe$_{20}$ structure (see Fig. 2) because the direction of $J_S$ is reversed in the Ni$_{80}$Fe$_{20}$/Nb inverted structure, providing an additional evidence of the spin-Hall voltages from the Nb [8-10].

The vanishing of the iSHE voltage for the patterned samples ($d_y \leq 30\,nm$) well below $T_c$ suggests the rapid decay of $\lambda_Q$ of Nb as $T_c$ is crossed. These results are in contrast to a previous observation of the giant iSHE induced by electrical spin injection from Ni$_{80}$Fe$_{20}$ through Cu into superconducting NbN ($d_y \approx 400\,nm$) far below $T_c$ [25]. However, a recent report on the iSHE voltage produced by the spin Seebeck effect in a YIG/NbN bilayer measurable only in a limited $T$ range right below $T_c$ [26] is more consistent with our findings. We note further that $\lambda_Q$ is typically larger than the superconducting coherence length $\xi_{SC}$ and comparable to $l_{sd}$ at a lower $T$ in the experiments performed to date [47-49]; thus it appears that a shorter $\lambda_Q$ is predicted in NbN relative to Nb [25,34]. The exact origin of the observed differences between experiments is not yet clear although different materials, device geometry, contact property, spin injection method, and spin-orbit coupling mechanism will undoubtably have influence, requiring further investigation. A natural starting point for the further work is to develop a spin Hall device [63] that works reliably in both the normal and (deep into) the superconducting states with a reasonable driving power density, as proposed here.

## IV. CONCLUSIONS

We experimentally estimated the RT values of $\theta_{SH}$, $l_{sd}$, and $\tau_{sf}$ of Nb films from spin-pumping-induced iSHE measurements in Nb/Ni$_{80}$Fe$_{20}$ bilayers by varying $t_{Nb}$, comparing to a Ni$_{80}$Fe$_{20}$/Pt reference sample, and measuring an out-of-plane angular dependence. We also studied the iSHE as a function of $T$ for different $t_{Nb}$. Above $T_c$ of Nb, a clear $t_{Nb}$-dependent $T$ evolution of the iSHE is observed whereas below $T_c$, the iSHE voltage drops abruptly and becomes undetectable at a lower $T$. This can be understood in terms of the strong decay of $\lambda_Q$ across $T_c$ of the Nb, as supported by the additional investigation of the iSHE in a different sample geometry along with model calculation. Our results suggest that the QP charge-imbalance relaxation length (of superconducting Nb) is shorter than hitherto assumed and needs to be considered in the development of new spin-pumping and spin-torque FMR devices [63] that aim to utilize QP spin-to-charge conversion and vice versa, respectively.

## ACKNOWLEDGMENTS

This work was supported by EPSRC Programme Grant EP/N017242/1.

# FIGURE CAPTIONS

FIG. 1. (Color online) (a) Sketch of the experimental setup used to dynamically inject a pure spin current $J_s$ and electrically detect a (transverse) charge current $J_c$ converted via inverse spin Hall effect in a Nb/Ni$_{80}$Fe$_{20}$ bilayer. (b),(c) Spatial profile of the inverse spin Hall voltage $V_{iSHE}^{(Q)}$ induced by spin pumping in a Nb/Ni$_{80}$Fe$_{20}$ bilayer above and below the superconducting transition temperature $T_c$ of Nb. In Fig. 1(c), $\exp[-d_y/\lambda_Q]$ describes the spatial decay of the charge-imbalance effect, where $\lambda_Q$ is the quasiparticle charge-imbalance relaxation length and $d_y$ is the distance between the inside edges of the precessing Ni$_{80}$Fe$_{20}$ and the voltage contact. The wine dashed line represents the active regime of ferromagnetic resonance in the Ni$_{80}$Fe$_{20}$. Note that the lateral dimension of the sample is much larger than the spin diffusion length of Nb.

FIG. 2. (Color online) (a)-(c) Ferromagnetic resonance absorption (top panel) and DC voltage measurements (bottom panel) vs. external magnetic field $\mu_0 H$ (along the $x$-axis) for the Nb/Ni$_{80}$Fe$_{20}$ sample with three different Nb thicknesses $t_{Nb}$ (7.5, 30, and 60 nm) at 300 K. In these measurements, the MW frequency was fixed at 5 GHz and the MW power at the CPW at ~100 mW. The solid lines are fits to Lorentzians [Eq. (1)]. (d),(e) Typical example of the $P_{MW}$ dependence of symmetric Lorentzian $V_{sym}$, extracted from fitting Eq. (1) to the data of $t_{Nb} = 7.5$ nm [Fig. 2(d)]. The black solid line is a linear fit. (f) The data shown is similar to that in Figs. 2(a)-(c) but now for the Pt(5 nm)/Ni$_{80}$Fe$_{20}$ reference sample.

FIG. 3. (Color online) (a) Effective Gilbert damping $\alpha$ as a function of Nb thickness $t_{Nb}$. The inset summarizes the effective saturation magnetization $\mu_0 M_{eff}$ for each $t_{Nb}$. These were deduced from the MW frequency $f$ dependence of FMR spectra (see Ref. [28]). Fitting Eq. (2) to the data (blue solid line) yields $g_r^{\uparrow\downarrow} = 16 \pm 3$ nm$^{-2}$ and $l_{sd}^{SC} = 35 \pm 2$ nm at 300 K. (b) Symmetric Lorentzian of DC voltage $V_{sym}$ as a function of $t_{Nb}$. The red solid line represents the room temperature values obtained from Eq. (4) for $\theta_{SH} \approx -0.001$ and $l_{sd}^{SC} \approx 30$ nm in Nb films.

FIG. 4. (Color online) (a) Out-of-plane magnetic-field-angle dependence of DC voltage $V - V_0$ obtained from the Nb(30 nm)/Ni$_{80}$Fe$_{20}$ sample, taken at a fixed MW frequency $f$ of 10 GHz and MW power $P_{MW}$ of ~100 mW. The inset illustrates schematically the measurement scheme. $\theta_H$ ($\phi_M$) is the angle of external magnetic field (magnetization precession axis of FM) to the $x$-axis. (b) Top panel. $\theta_H$ dependence of the resonance field. The upper inset displays the calculated $\phi_M$ as a function of $\theta_H$ using Eq. (5). (b) Bottom panel. $\theta_H$ dependence of the symmetric Lorentzian $V_{sym}$, extracted from fitting Eq. (1) to the data of Fig. 4(a). The measured $V_{sym}(\theta_H)$ is fairly reproduced by Eq. (6) with the spin lifetime $\tau_{sf}$ of the order of a few ps (lower inset). For comparison, the calculated $V_{iSHE}(\theta_H)$ using Eq. (6) with $\tau_{sf} \ll 1/\omega_L$ (red solid line), $\tau_{sf} = 1/\omega_L$ (black solid line), and $\tau_{sf} \gg 1/\omega_L$ (blue solid line) are also shown.

FIG. 5. (Color online) (a)-(c) Temperature $T$ evolution of DC voltage $V-V_0$ for the Nb/Ni$_{80}$Fe$_{20}$ samples with three different Nb thicknesses $t_{Nb}$ of 7.5, 30, and 60 nm, taken at a fixed MW frequency $f$ of 5 GHz. Note that for more quantification, the $V-V_0$ value is normalized by the MW power $P_{MW}$. (d) $T$ dependence of the normalized symmetric Lorentzian $V_{sym}/P_{MW}$, extracted from fitting Eq. (1) to the data of Fig. 5(a), for $t_{Nb}$ = 7.5, 30, and 60 nm. The inset shows the normalized resistance $R/R_{300\,K}$ vs. $T$ plot for bare Nb films.

FIG. 6. (Color online) (a) Calculated values of the quasiparticle (QP) spin susceptibility $\chi_S^0(T)$ divided by the QP population $2f_0(\Delta)$ (top panel), the effective spin transport length $l_Q^*$ (middle panel), and the spin current density $j_s^Q$ at a Nb/Ni$_{80}$Fe$_{20}$ interface (bottom panel) using Eqs. (S4)-(S6), respectively, across the superconducting transition temperature $T_c$ of Nb. The green and pink curves represent respectively the superconducting Nb/Ni$_{80}$Fe$_{20}$ samples with the Nb thicknesses $t_{Nb}$ of 30 and 60 nm. (b),(c) Calculated DC voltage $V_{iSHE}^Q$ expected from the QP-mediated inverse spin Hall effect, using Eqs. (S3)-(S6), for $t_{Nb}$ = 30 (top panel) and 60 nm (bottom panel) across their $T_c$. Each inset presents the dependence of $V_{iSHE}^Q$ on the active width of the precession Ni$_{80}$Fe$_{20}$, $w_y$, above and well below $T_c$. Figures 6(b) and 6(c) indicate respectively the side jump and skew scattering contributions. Note that a larger increase of $V_{iSHE}^Q$ at a lower $T$ in the skew scattering case relative to that in the side jump reflects its strong $T$ dependence, $\propto \chi_S^0(T)/2f_0(\Delta)$ [see Fig. 6(a), top panel] [50,51].

FIG. 7. (Color online) (a) Schematic of the single-stripe-patterned sample, comprising an un-etched Ni$_{80}$Fe$_{20}$/superconducting Nb bilayer at the middle and etched non-superconducting Nb leads (< 7.5 nm) on the lateral sides of the bilayer. (b) Normalized resistance $R/R_N$ vs. temperature $T$ plots measured at the un-etched Ni$_{80}$Fe$_{20}$/Nb bilayer (closed green symbol) and at the etched Nb lead (open green symbol) using a four-point current-voltage method without MW excitation. (c) Scanning electron microscope images of the patterned sample. (d)-(g) The data shown is similar to that in Fig. 6 but now for the patterned samples with the Ni$_{80}$Fe$_{20}$ spin source width $w_y$ of 150 and 500 μm.

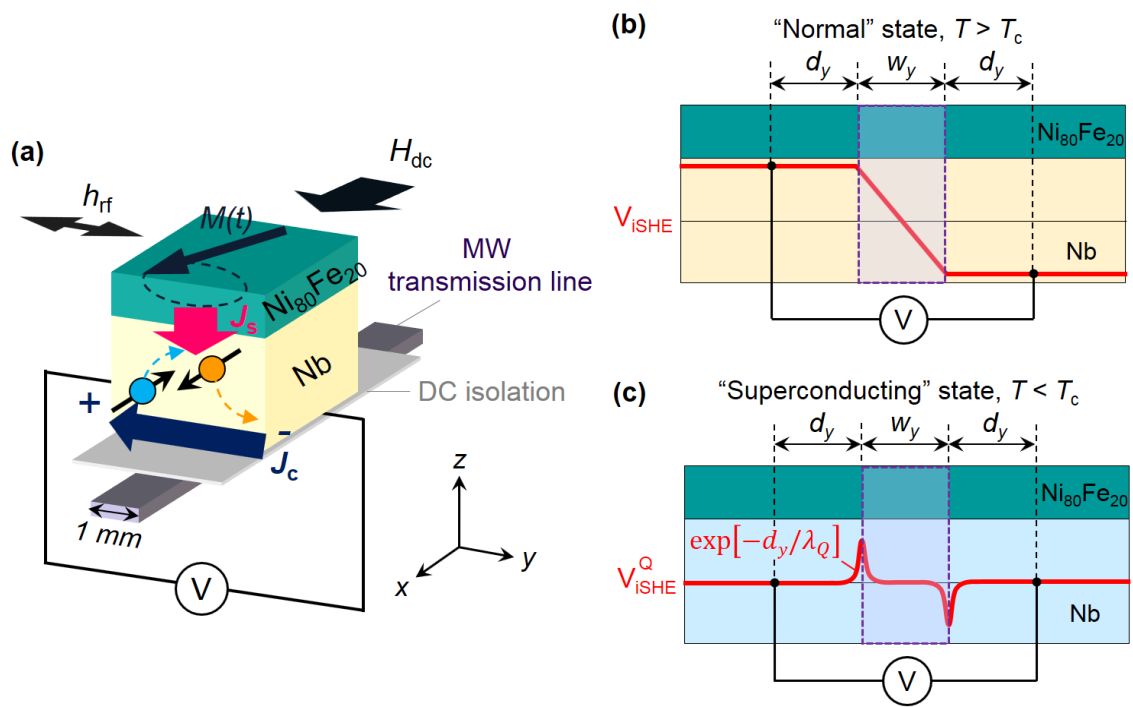

FIG. 1

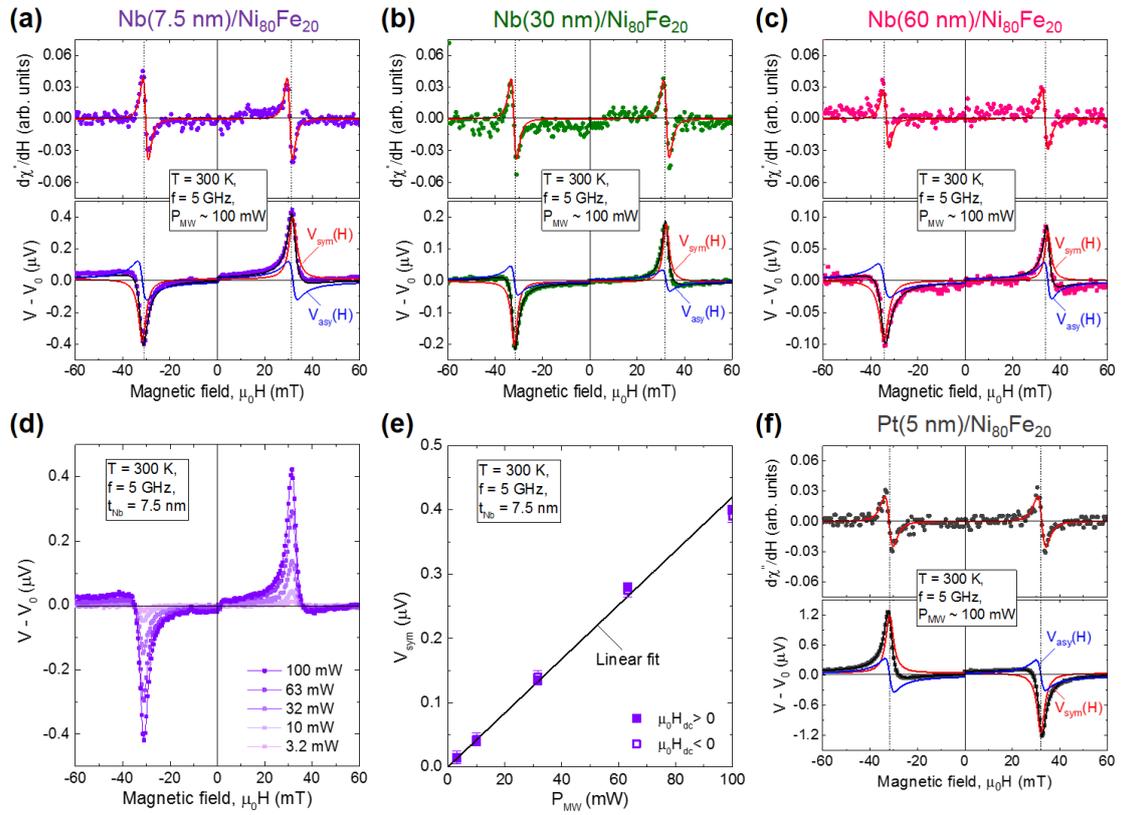

FIG. 2

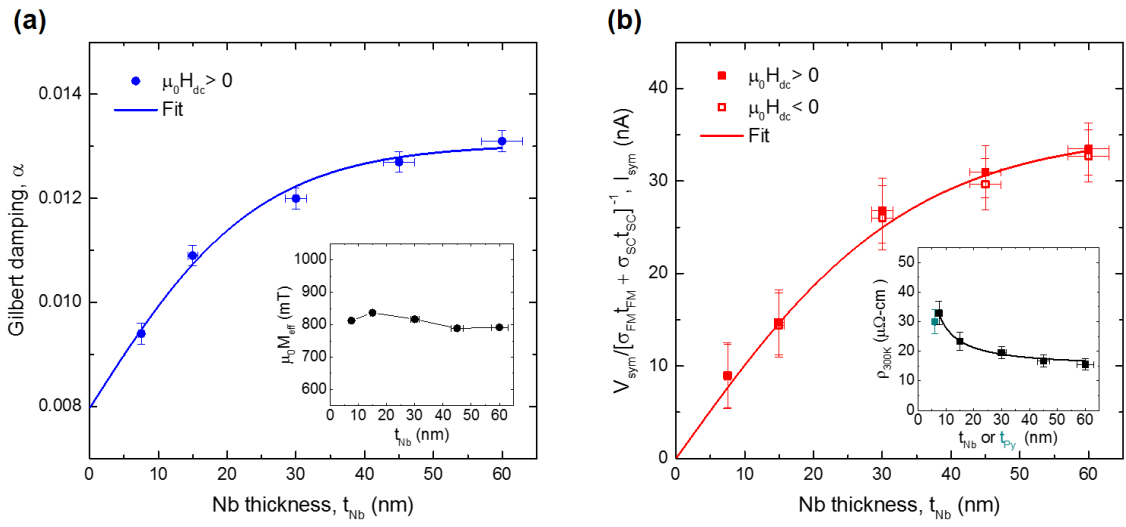

FIG. 3

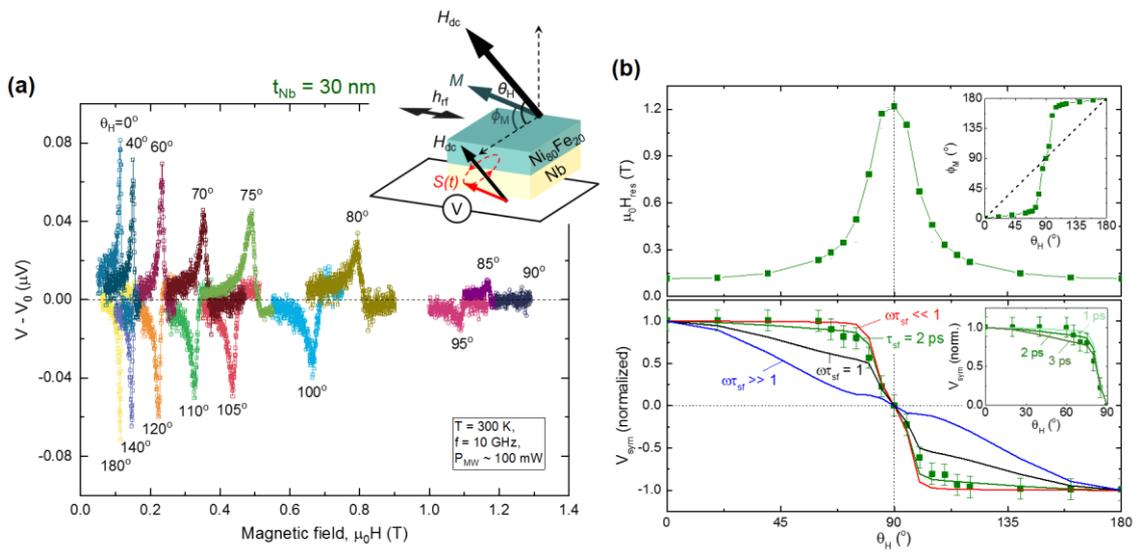

FIG. 4

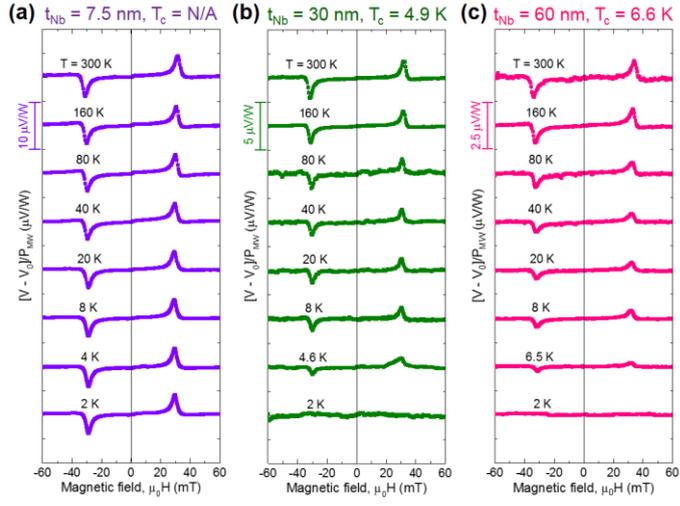
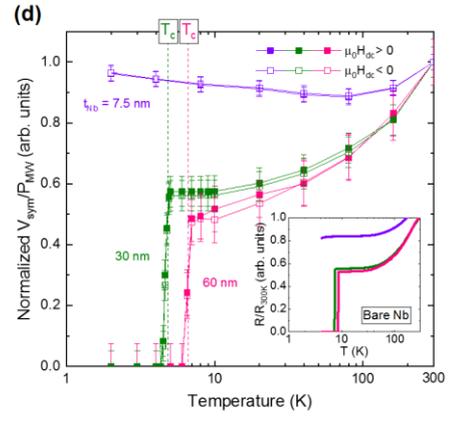

FIG. 5

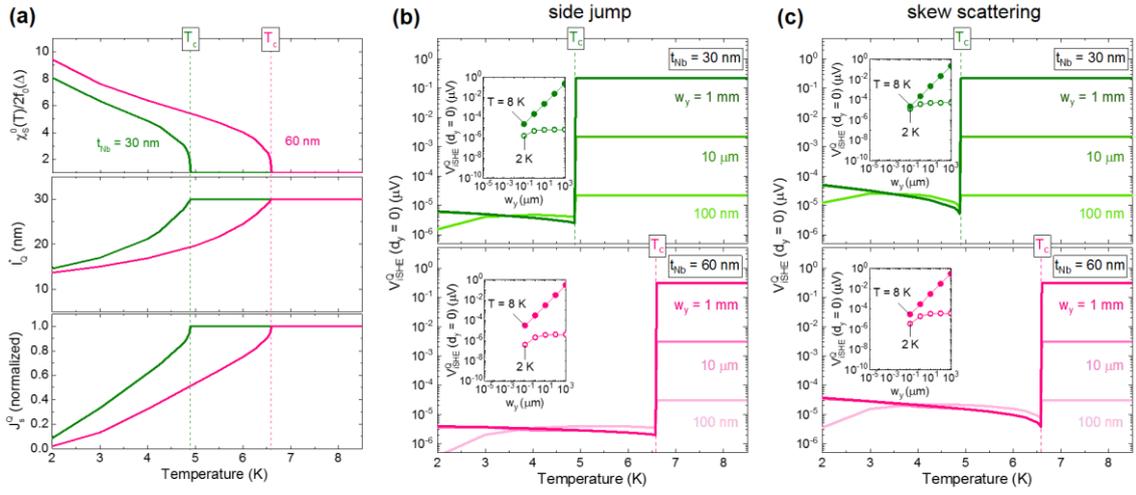

FIG. 6

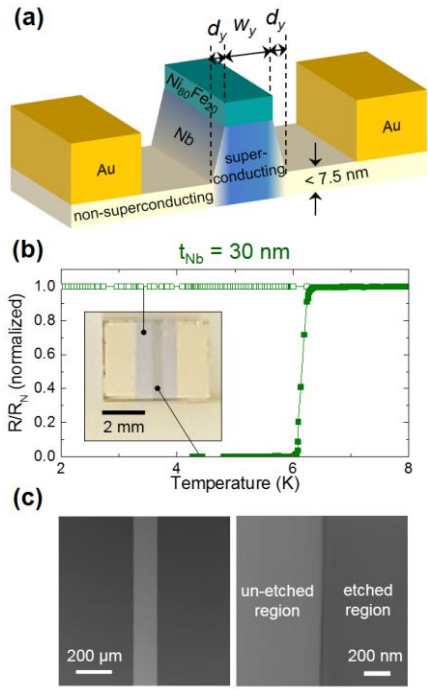

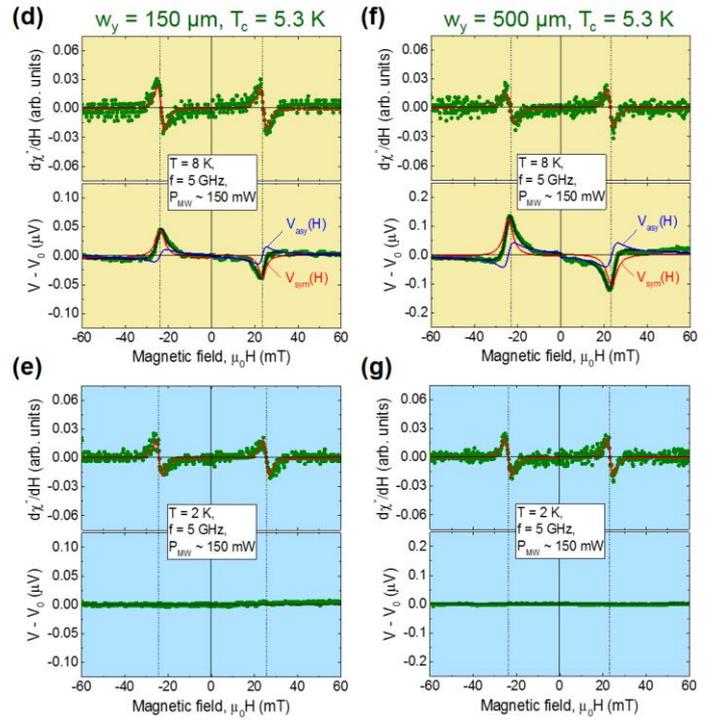

FIG. 7

# Supplementary Material

## Section S1. Effect of MW power on the superconductivity of Nb.

The effect of MW power on the superconducting property of Nb in terms of unintentional heating was investigated by measuring the 2-terminal resistance $R$ vs. $T$ curves for the Nb(30 nm)/Ni$_{80}$Fe$_{20}$ sample with varying $P_{MW}$ [Fig. S1(a)]. As $P_{MW}$ increases, there is a clear shift of the superconducting transition to a lower $T$, as summarized in the inset of Fig. S1(a).

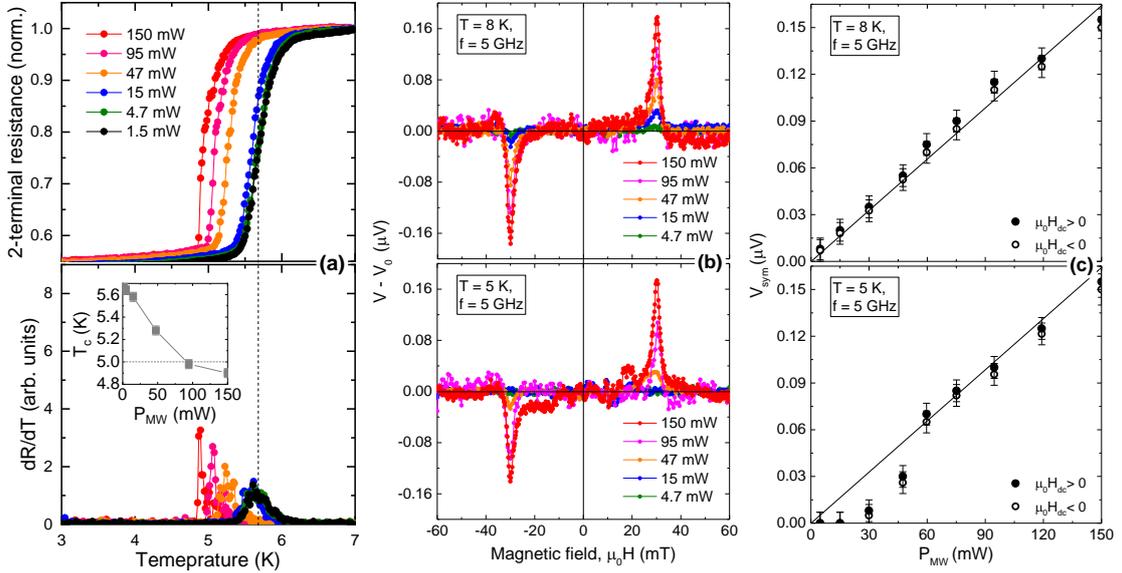

FIG. S1. (a) 2-terminal resistance vs. temperature $T$ plots acquired from the Nb(30 nm)/Ni$_{80}$Fe$_{20}$ sample with varying the MW power $P_{MW}$ (top panel). From $T$ derivative of $R$ (bottom panel), d$R$/d$T$, the superconducting transition temperature $T_c$ was determined as the $T$ value that exhibits the maximum of d$R$/d$T$. The inset summarizes the $P_{MW}$ dependence of $T_c$. The vertical dashed line represents the $T_c$ value (~5.7 K) obtained from the same sample in a separate liquid helium dewar using a four-point current-voltage method without MW excitation. (b) $P_{MW}$ dependence of DC voltages taken above (top panel) and immediately below (bottom panel) $T_c$. (c) Corresponding $P_{MW}$ dependence of the symmetric Lorentzian $V_{sym}$, extracted from fitting Eq. (1) to the data of Fig. S1(b). The black solid lines are linear fits.

To further check the heating effect, we also measured the $P_{MW}$ dependence of DC voltages above and immediately below $T_c$ [Fig. S1(b)]. By comparing the $V_{sym}$ vs. $P_{MW}$ plots in Fig. S1(c), one can see that $V_{sym}$ obtained at 5 K deviates from the linear scaling and diminishes rapidly for $P_{MW} < 60$ mW, where the local/actual $T$ is below the superconducting transition of Nb [see Fig. S1(a)]. Nevertheless, the finite voltage signals for $P_{MW} < 60$ mW implies that the charge-imbalance effect around $T_c$ is *non-ignorable*, as expected from the model calculation (see Fig. 6) and also from Refs. [47,48].

## Section S2. MW frequency dependence of FMR spectra.

The MW frequency $f$ dependences of $\mu_0 H_{res}$ and $\mu_0 \Delta H$ for the Nb/Ni$_{80}$Fe$_{20}$ (normal structure) samples are respectively summarized in Figs. S2(a) and S2(b). The dispersion relation of $\mu_0 H_{res}$ with $f$ can be described by Kittel's formula:

$$f = \frac{\gamma}{2\pi} \sqrt{[\mu_0 (H_{res} + M_{eff}) \cdot \mu_0 H_{res}]}, \quad \text{(S1)}$$

The values of $\mu_0 M_{eff}$ determined from Fig. S2(a) using Eq. (S1) are in the range of $790 - 840$ mT. In Fig. S2(b) where $\mu_0 \Delta H$ scales linearly with $f$ for all cases, we can calculate the Gilbert-type damping constant $\alpha$ using the following equation:

$$\mu_0 \Delta H(f) = \mu_0 \Delta H_0 + \frac{4\pi \alpha f}{\sqrt{3} \gamma} \quad \text{(S2)}$$

with $\mu_0 \Delta H_0$ is the zero-frequency line broadening due to long-range magnetic inhomogeneities [S1] in the FM. All of the samples have small $\mu_0 \Delta H \leq |0.4$ mT|, meaning the high quality of the samples and the absence of two-magnon scattering. We note that the clear enhancement of $\alpha$ with $t_{Nb}$ from $9.4 \times 10^{-3}$ to $13.1 \times 10^{-3}$ in Fig. S2(b) is the indicative of spin pumping effect in the Nb layers [8-10].

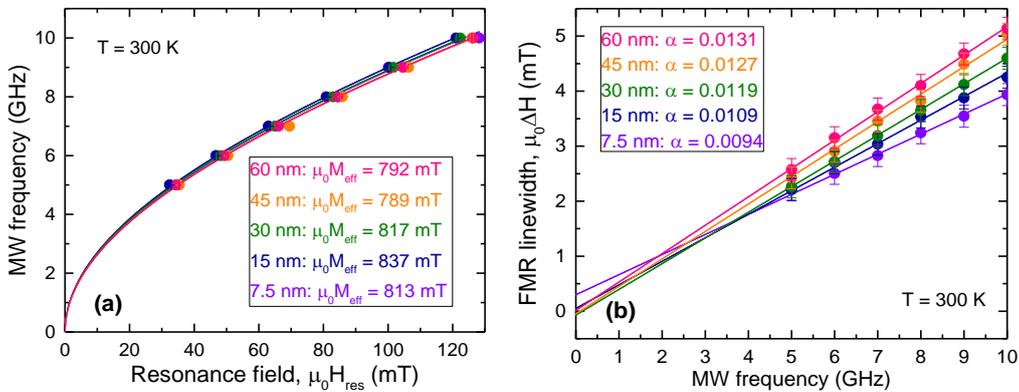

FIG. S2. (a) Microwave frequency $f$ vs. resonance magnetic field $\mu_0 H_{res}$. The solid lines

are fits to estimate the effective saturation magnetization $\mu_0 M_{eff}$ via Kittel's formula [Eq. (S1)]. (b) FMR linewidth $\mu_0 \Delta H$ as a function of $f$. The solid lines are fitting curves to deduce the Gilbert damping constant $\alpha$ using Eq. (S2).

## Section S3. Proposal of the device geometry for amplifying QP spin-Hall voltages.

In the main text, we proposed a device geometry to amplify the QP-mediated spin-Hall voltage, namely an array of densely-packed FM stripes with a periodicity $\Lambda$ of the order of $\lambda_Q$ [Fig. S3(a)]. In such a geometry, one can greatly increase the active volume of QP charge imbalance and thereby the total amplitude of spin-Hall voltage for a given constant $P_{MW}$. Figure S3(b) presents the calculated $V_{iSHE}^Q$ for the proposed device using Eqs. (8)-(11). Note that in this calculation, we assumed that $d_y = 0$ and $w_y = d_s = \Lambda/2$, and thus the estimated value should be considered as the upper limit of $V_{iSHE}^Q$. Notably, from the measured value of $V_{sym}=50-150$ nV (see Fig. 6), we have $V_{iSHE}^Q$ of the order of $10-100$ nV [Fig. S3(b)], which can be *measurable* well below $T_c$ regardless of details of the QP spin-Hall mechanism [47,48]. Hence, we believe that the proposed spin-pumping device can be employed not only to probe the QP-mediated iSHE [S2] but also provide a new spin-torque FMR device [S3,S4] utilizing its reciprocal effect.

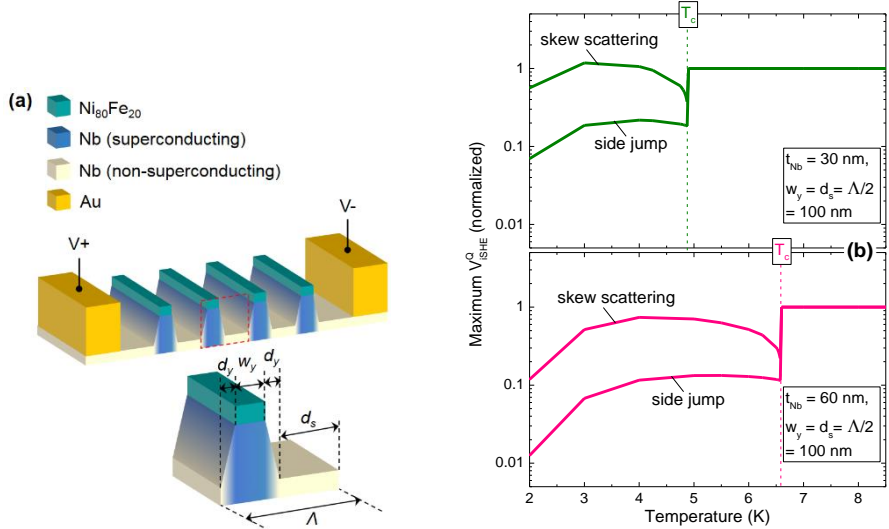

FIG. S3. (a) Schematic of the proposed spin-pumping device for amplifying the quasiparticle-mediated spin-Hall voltage $V_{iSHE}^Q$: An array of densely-packed ferromagnet stripes with a periodicity $\Lambda$ of the order of the superconducting coherence length $\lambda_Q$. (b) Calculated values of $V_{iSHE}^Q$ for the proposed device with two different Nb thicknesses $t_{Nb}$

of 30 (top panel) and 60 nm (bottom panel), using Eqs. (8)-(11). In this calculation, we assumed that $d_y = 0$ and $w_y = d_s = \Lambda/2 = 100$ nm for simplicity. So the estimated value should be considered as the upper limit of $V_{iSHE}^Q$.

### Section S4. Control experiment on a Nb/Ni$_{80}$Fe$_{20}$/Nb symmetric structure.

It was shown in the main text that the iSHE in Nb layers can be responsible for the observed transverse DC voltages in our experimental setup by showing 1) Hanle spin precession under an oblique magnetic field (see Fig. 4) and 2) sign inversion of the voltages for the inverted structure (see Fig. 7). In this section, we further confirm that by performing the control experiment on a Nb(30 nm)/Ni$_{80}$Fe$_{20}$(6 nm)/Nb(30 nm) symmetric structure [Fig. S4(a)]. As shown in Fig. S4(b), the symmetric Lorentzian of DC voltage $V_{sym}$ is significantly reduced by one order of magnitude compared to asymmetric structures [see Figs. 2(a)-2(c)]. This is because two charge currents ($J_c^1$ and $J_c^2$) in opposite directions [Fig. S4(a)], converted via the iSHE from the pumped spin currents ($J_s^1$ and $J_s^2$) in top and bottom Nb layers respectively, cancel each other out [8-10]. Note that a non-vanishing $V_{sym}$ (~10 nV) in the symmetric structure might be due to incomplete calculation of $J_c^1$ and $J_c^2$ as the interfaces of Ni$_{80}$Fe$_{20}$ grown on Nb and Nb grown on Ni$_{80}$Fe$_{20}$ are likely to be different [S5]. Consequently, we believe that the control experiment provides a decisive evidence for the spin-Hall voltages originating from the Nb.

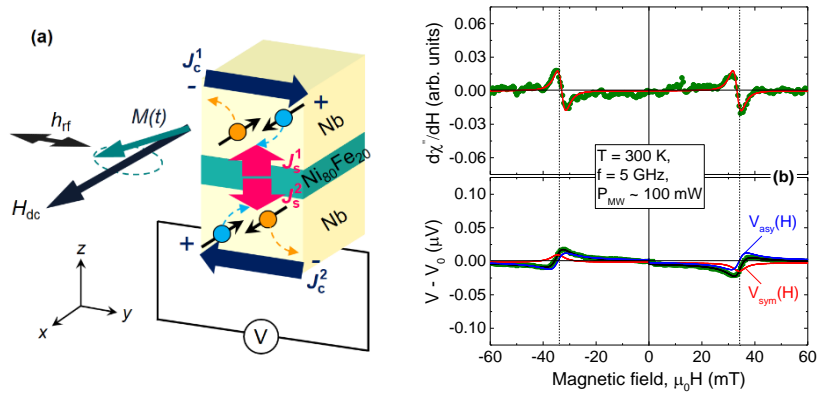

FIG. S4. (a) Sketch of the control experiment on a Nb(30 nm)/Ni$_{80}$Fe$_{20}$(6 nm)/Nb(30 nm) symmetric structure. (b) Ferromagnetic resonance absorption (top panel) and DC voltage measurements (bottom panel) vs. external magnetic field $\mu_0 H$ (along the $x$-axis) for the Nb/Ni$_{80}$Fe$_{20}$/Nb sample at 300 K. In these measurements, the MW frequency was fixed at 5 GHz and the MW power at the CPW at ~100 mW.

A monotonic decay of iSH voltages in the un-patterned and stripe-patterned samples across $T_c$ (Fig. S5) confirms the absence of the superconducting coherence effect [44-46] in our system (*i.e. metallic/conducting* FM/SC bilayers).

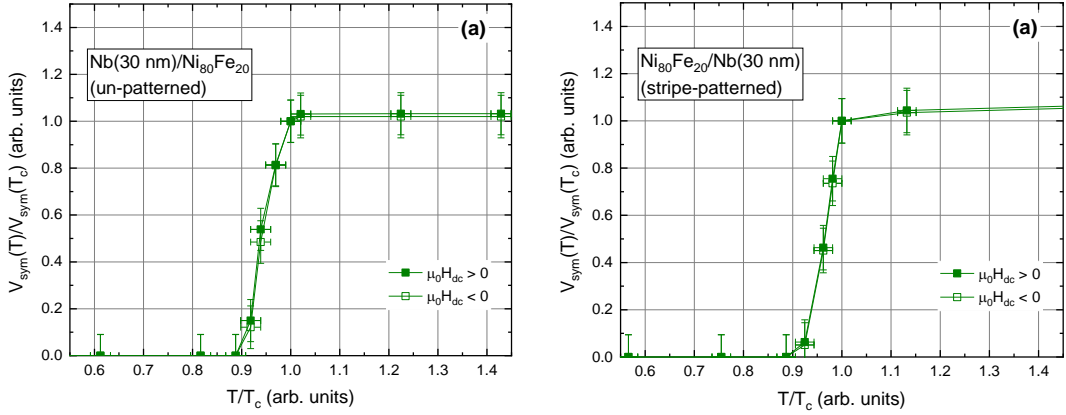

FIG. S5. Normalized symmetric Lorentzian $V_{sym}(T)/V_{sym}(T_c)$ as a function of normalized temperature $T/T_c$ for (a) the un-patterned Nb(30 nm)/Ni$_{80}$Fe$_{20}$ and (b) the stripe-patterned Ni$_{80}$Fe$_{20}$/Nb(30 nm) ($w_y \approx 500$ µm) samples.

**Section S6. Theoretical description of QP-mediated spin-Hall voltages**

The transverse DC voltage $V_{iSHE}^Q$ expected from QP-mediated iSHE in the superconducting Nb layer is following. When $t_{SC} < \lambda_Q$ and $t_{SC} \sim l_{sd}$, as relevant to our geometry [see Fig. 1(c)],

$$V_{iSHE}^Q = \left( \frac{R_{FM} R_{SC}^Q}{R_{FM} + R_{SC}^Q} \right) \cdot I_c^Q$$

$$\approx \left[ \frac{w_y}{\sigma_{FM} t_{FM} + \sigma_{SC}^Q t_{SC} \cdot \left( \frac{w_y/2\lambda_Q}{\tanh(w_y/2\lambda_Q)} \right)} \right] \cdot \theta_{SH}^Q l_Q^* \cdot \tanh\left( \frac{t_{SC}}{2 l_Q^*} \right) \cdot j_s^Q \cdot \exp\left[ -\frac{d_y}{\lambda_Q} \right], \quad \text{(S3)}$$

$$\theta_{SH}^Q = \theta_{SH}^{SJ} + [\chi_S^0(T)/2 f_0(\Delta)] \cdot \theta_{SH}^{SS}, \quad \text{(S4)}$$

$$l_Q^* \approx \sqrt{D_S \cdot \left( \frac{1}{\tau_{AR}} + \frac{1}{\tau_{sf}} \right)^{-1}}, \quad \text{(S5)}$$

$$j_s^Q \approx g_r^{\uparrow\downarrow} \cdot \left[ 1 + \frac{g_r^{\uparrow\downarrow} \mathcal{R}_{SC}^Q}{\tanh\left( \frac{t_{SC}}{l_Q^*} \right)} \right]^{-1} \cdot \left( \frac{\hbar}{8\pi} \right) \cdot \left( \frac{\mu_0 h_{rf} \gamma}{\alpha} \right)^2 \cdot \left[ \frac{\mu_0 M_{eff} \gamma + \sqrt{(\mu_0 M_{eff} \gamma)^2 + 16(\pi f)^2}}{(\mu_0 M_{eff} \gamma)^2 + 16(\pi f)^2} \right] \cdot \left( \frac{2e}{\hbar} \right),$$

here $R_{SC}^Q \approx \left[(2\lambda_Q/w_y)\cdot\tanh\left(w_y/2\lambda_Q\right)\right]\cdot\left(\rho_{SC}^Q w_y/t_{SC}w_x\right)$ is the effective QP resistance [50,51]. Note that $\left[(2\lambda_Q/w_y)\cdot\tanh\left(w_y/2\lambda_Q\right)\right]$ represents an estimate for the volume of the charge imbalance contributing to the Nb resistance below $T_c$ [see Fig. 1(c)] [50,51]. $\rho_{SC}^Q \approx \rho_0/[2f_0(\Delta)]$ is the QP resistivity [25] and $\rho_0$ is the residual resistivity of the Nb layer (7−8 μΩ-cm) immediately above $T_c$ [27]. $I_c^Q$ is the QP current and $\mathcal{R}_{SC}^Q \equiv \rho_{SC}^Q l_Q^* e^2/2\pi\hbar$ is the spin resistance of QP. Based on the previous theoretical framework [50,51], we speculate that the QP spin Hall angle $\theta_{SH}^Q$ is given by two extrinsic components: the side jump $\theta_{SH}^{SJ}$ [54] and the skew scattering $\theta_{SH}^{SS}$ [55]. $\chi_S^0(T) = 2\int_\Delta^\infty \left(E/\sqrt{E^2-\Delta^2}\right)\cdot\left[-\partial f_0(E)/\partial E\right]dE$ is the normalized spin susceptibility of the QP [47,48]. It is notable that the side jump contribution is $T$-independent while the skew scattering is gradually enhanced as $T$ is reduced [Fig. 6(a), top panel]. $l_Q^*$ is the effective spin transport length considering the conversion time $\tau_{AR}$ of QPs into singlet Cooper pairs by Andreev reflection in addition to their $\tau_{sf}$ [20]. $D_S = [2f_0(\Delta)/\chi_S^0(T)]D$ is the spin diffusion coefficient of the QPs [50,51]. The postfactor $\exp\left[-d_y/\lambda_Q\right]$ in Eq. (8) represents the spatial decay of the charge imbalance effect, where $d_y$ is the distance between the inside edges of the precessing FM and the voltage contact [see Fig. 1(c)]. In this calculation, we assumed that $\tau_{sf}$ and $g_r^{\uparrow\downarrow}$ do not change significantly on entry to the superconducting state for simplicity. According to recent studies [44,45], a coherence effect of superconductivity can enhance the energy-dependent spin-flip scattering and thus $\tau_{sf}$ is expected to exhibit a non-monotonic $T$ dependence immediately below $T_c$, when the $E$ interval of QPs (order of $k_BT$) is comparable to the superconducting gap $\Delta(T)$. However for a metallic/conducting FM in direct contact with SC [27,46], $\Delta$ is significantly suppressed at the FM/SC interface due to the (inverse) proximity effect of the FM, leading to the vanishing of the energy-dependent spin-flip scattering associated with the superconducting coherence peak [44,45]. The energy scale of dynamically-driven spin-polarized QPs by FMR excitation is given by $\theta c\cdot(hf)$, where $\theta c$ is the precession cone angle [S2] and $h$ is Planck's constant. For small-angle precession (a few degrees) with $f$ = 5 GHz [27], the relevant energy scale (< μeV) is approximately 3 orders of magnitude smaller than the superconducting gap of Nb. Thus, we ignored the energy dependence of dynamic exchange coupling $g_r^{\uparrow\downarrow}$ [34,44].